\documentclass[
aip,
jcp,
reprint,
citeautoscript,
superscriptaddress
]{revtex4-2}  % for review and submission
\usepackage{bm}        % for math
\usepackage{amssymb}   % for math
\usepackage{amsmath}
\usepackage{mathtools}
\usepackage{physics}
\usepackage{hyperref}
\usepackage[normalem]{ulem}
\hypersetup{
	colorlinks=true,
	linkcolor=blue,
	filecolor=blue,
	citecolor = black,      
	urlcolor=blue,
}

\usepackage{xcolor}
\definecolor{amber}{rgb}{1,0.49,0}

\newcommand{\editor}[2]{%
	\expandafter\newcommand\csname #1note\endcsname[1]{%
		\textcolor{#2}{(\textbf{#1:} ##1)}}%
	\expandafter\newcommand\csname #1\endcsname[1]{%
		\textcolor{#2}{##1}}%
	\expandafter\newcommand\csname #1cancel\endcsname[1]{%
		\textcolor{#2}{\sout{##1}}}%
	\expandafter\newcommand\csname #1change\endcsname[2]{%
		\textcolor{#2}{\sout{##1} ##2}}%
	\expandafter\newcommand\csname #1prov\endcsname[2]{%
		\textcolor{#2}{[(##1) (##2)]}}%
	\newenvironment{#1text}{\color{#2}}{\color{black}}
}

\editor{SB}{amber}
\editor{ED}{purple}
\editor{MGI}{green}
\editor{CAVEAT}{red}
\editor{resub}{cyan}

\renewcommand{\[}{\begin{equation}}
	\renewcommand{\]}{\end{equation}}
\renewcommand{\(}{\begin{equation*}}
	\renewcommand{\)}{\end{equation*}}

\newcommand{\kk}{{\bm k }}
\newcommand{\RR}{{\bm R }}
\newcommand{\rr}{{\bm r }}

\newcommand{\hjj}{\hat{{\bm \jmath}}}

\newcommand{\pprime}{{\prime\prime}}

\usepackage{orcidlink}

\begin{document}

\title{Heat conductivity from energy-density fluctuations}
	
\author{Enrico Drigo\, \orcidlink{0000-0002-1797-2987}}
\email{endrigo@sissa.it}
\affiliation{SISSA – Scuola Internazionale Superiore di Studi Avanzati, Trieste, Italy}
\author{Maria Grazia Izzo}
\affiliation{SISSA – Scuola Internazionale Superiore di Studi Avanzati, Trieste, Italy}
\author{Stefano Baroni\,\orcidlink{0000-0002-3508-6663}}
\affiliation{SISSA – Scuola Internazionale Superiore di Studi Avanzati, Trieste, Italy}
\affiliation{CNR-IOM DEMOCRITOS, SISSA, Trieste, Italy}
	
\date{\today}
\begin{abstract}
We present a method, based on the classical Green-Kubo theory of linear response, to compute the heat conductivity of extended systems, leveraging energy-density, rather than energy-current, fluctuations, thus avoiding the need to devise an analytical expression for the macroscopic energy flux. The implementation of this method requires the evaluation of the long-wavelength and low-frequency limits of a suitably defined correlation function, which we perform using a combination of recently-introduced cepstral-analysis and Bayesian extrapolation techniques. Our methodology is demonstrated against standard current-based Green-Kubo results for liquid argon and water, and solid amorphous Silica, and compared with a recently proposed similar technique, which utilizes mass-density, instead of energy-density, fluctuations.
\end{abstract}

\maketitle

\section{Introduction}
Heat transport is one of the most fundamental off-equilibrium processes occurring in nature, whose understanding sinks its roots into the classic works of Onsager in the thirties \cite{onsagerI1931,*onsagerII1931} and of Green and Kubo (GK) \cite{green1952,*green1954,kubo1957a,*kubo1957b} in the fifties. In spite of the strong theoretical foundation that these works provide to the numerical simulation of transport phenomena in condensed matter, until recently widespread misconceptions had hindered the application of standard equilibrium simulation techniques to the numerical modeling of heat transport when atomic forces are derived from quantum-mechanical \emph{ab initio} methods \cite{stackhouse2010}. While the crux of the difficulties was originally ascribed to the impossibility of assigning a well-defined value to the atomic energies that enter the definition of the macroscopic energy current---when atomic forces are derived from quantum mechanical methods---it was later realized that a unique partition of the total energy into local contributions is not possible even when atomic forces are derived from classical potentials. It was then demonstrated that, although the macroscopic current corresponding to different local partitions of the total energy depends on the specific partition being considered, the value of the resulting heat conductivity does not. This independence of the heat conductivity of the local representation of the total energies was dubbed \emph{gauge invariance} of transport coefficients \cite{marcolongo2016,Ercole2016,grasselli2021, marcolongo_2020} and has far-reaching consequences in other branches of transport theory as well, such as, \emph{e.g.}, charge transport in ionic conductors \cite{Grasselli2019,Pegolo2020}. 

Although gauge invariance provides a general framework to compute heat transport coefficients for general atomic forces derived from either electronic-structure theory or arbitrary---possibly machine-trained---inter-atomic potentials, the problem still remains that the expression for the macroscopic energy current for general classical force fields or quantum density functionals may be difficult to compute analytically and to implement numerically. For this reason, it would be desirable to devise a method to compute the heat conductivity not relying on the macroscopic energy current, which is the standard ingredient of the GK formula for the transport coefficient, but rather on the energy density, which---albeit not uniquely defined because of gauge freedom---is easier to compute. A successful attempt in this direction has been recently made by Cheng and Frenkel (CF) \cite{cheng2020} who have shown  how to evaluate the heat conductivity of simple fluids, $\kappa$,  by analyzing the wave-vector dependence of the widths of the Rayleigh (R) and Brillouin (B) peaks, as estimated from equilibrium molecular dynamics (EMD) simulations:\cite{KADANOFF1963419,forster,hansen2013} 

\begin{align}
    \Gamma_R(k)&= D_T k^2, \label{eq:Rayleigh} \\
    \Gamma_B(k)&= \frac{1}{2} \left (D_\ell + R_{LP} D_T \right )k^2, \label{eq:Brillouin}
\end{align}
where $D_T=\frac{\kappa}{mnc_p}$ and $D_\ell$ are the heat and longitudinal diffusivities, respectively, $R_{LP}=\frac{c_p}{c_v}-1
%\ED{=\left(\frac{\partial P}{\partial T}\left.\right\vert_{V}\right)^2\frac{\partial V}{\partial P}\left.\right\vert_{T}}
$ is the (Landau-Planczek) ratio between the intensities of the Rayleigh and Brillouin peaks , which vanishes for incompressible fluids, $c_p$ and $c_v$ are the isobaric and isochoric specific heats, and $m$ and $n$ the molecular mass and average number density, respectively. The smallness of the Landau-Planczek ratio in the condensed phases ($R_{LP}<0.01$ in liquid water at ambient conditions),\cite{Cengel2011} however, makes the application of this method numerically demanding in liquids and practically prohibitive in solids.

In a pioneering, but not highly cited, 1994 paper, Palmer showed how the heat conductivity can be determined by a direct analysis of the energy-density fluctuations in EMD simulations.\cite{palmer1994} This method does not rely on the value of the Landau-Planczek ratio, and therefore can be equally applied to liquids and solids. However, its broad applicability is also plagued by the numerical and statistical problems related to the evaluation and long-wavelength extrapolation of the peak of the energy dynamical structure factor (see below for a precise definition) and, on a more conceptual level, by the intrinsic indeterminacy of the energy density at the molecular scale, discussed above.

Building on the work of Palmer,\cite{palmer1994} in this paper we first reexamine the concept of gauge invariance in the specific context of the energy-density approach to heat transport, and then we show that a combination of recent advances in data analysis and signal processing, based on Bayesian inference \cite{bishop2006} and cepstral analysis \cite{Ercole2017,sportran}, can be leveraged to overcome the numerical difficulties that have plagued the application of Palmer's approach. We argue that the proposed methodology naturally lends itself to a quantitative evaluation of the resulting statistical uncertainties and can be applied using simulation cells of moderate size and EMD runs of moderate length; we finally showcase it with three applications to the paradigmatic cases of liquid water and Argon, and solid amorphous Silica.

\section{Theory}

The standard GK expression for the heat conductivity, $\kappa$, reads \cite{KADANOFF1963419,forster,hansen2013}:

\begin{align}
    \kappa &= \frac{V}{k_B T^2} \int_0^\infty \langle\widehat J(t)\widehat J(0) \rangle dt \label{eq:GK} \\
           &=\frac{1}{T} \lim_{\omega\to 0}\lim_{\kk\to 0} \frac{\omega}{k^2}\widetilde\chi^\pprime(\kk,\omega). \label{eq:GK_chi}
\end{align}
In Eq. \eqref{eq:GK} $V$ and $T$ are the system's volume and temperature, $k_B$ the Boltzmann's constant, $\widehat J$ is any Cartesian component of the macroscopic average of the heat current density, $\widehat{\bm J} =\frac{1}{V} \int_V \hjj_q(\rr)d\rr$, a hat indicates an implicit dependence on phase-space variables, $\Gamma$---as in $\widehat J=J(\Gamma)$---and $\langle\cdot\rangle$ indicates an equilibrium average over the initial conditions of a molecular trajectory. The heat current density is defined as:\cite{KADANOFF1963419,forster,hansen2013}  $\hjj_q=\hjj_e-(p+e)\hjj_n/n$, where $\hjj_e$ and $\hjj_n$ are the energy and molecular-number density currents, respectively, and $p$, $e$ are the system's pressure and average energy, respectively. The explicit expressions of the energy and energy-current densities depend on the specific simulation framework (e.g. force-field vs. \emph{ab initio}), as well as on gauge fixing (see below). A detailed derivation of the relevant formulas is presented, e.g., in Ref. \onlinecite{Ercole2016}. 

In a one-component system, the macroscopic average of the molecular number current density is proportional to the total momentum, which, being conserved, can be assumed to vanish. As in this work we will only consider this class of systems, we will not make any difference between heat and energy densities, and we will therefore omit any suffixes aimed at distinguishing one from the other. In Eq. \eqref{eq:GK_chi} $\widetilde\chi^\pprime$ is the imaginary part of the heat-temperature susceptivity, which, according to the fluctuation-dissipation theorem, is proportional to the {energy dynamical structure factor} (EDSF), $\widetilde C(\kk,\omega)$:

\begin{align}
\widetilde\chi^\pprime(\kk,\omega)&=\frac{\omega}{2k_BT} \widetilde C(\kk,\omega), \label{eq:fluctuation-dissipation} \\
    \widetilde C(\kk,\omega)&=V \int_{-\infty}^\infty \left\langle \widehat{\widetilde e}(\kk,t) \widehat{\widetilde e}(-\kk,0)\right\rangle e^{i\omega t}dt, \label{eq:Ctilde}
\end{align}
where $\widehat{\widetilde e}(\kk)$ is the space Fourier transform of the energy density, $\widehat e(\rr)$. The equivalence between Eqs. \eqref{eq:GK} and \eqref{eq:GK_chi} follows from the fluctuation-dissipation theorem, Eq. \eqref{eq:fluctuation-dissipation}, and the continuity equation, which, in the frequency-wavenumber domain, reads $\omega\widehat{\widetilde e}(\kk,\omega)=\kk\cdot \widehat{\widetilde {\bm\jmath}}(\kk,\omega)$. Thanks to this relation, the density-density and current-current correlation functions can be expressed in terms of each other as: $\widetilde C_{\epsilon\epsilon}(\kk,\omega)=\frac{k^2}{\omega^2}\widetilde C_{jj}(\kk,\omega)$.

By combining the Onsager's regression hypothesis\cite{onsagerII1931,*onsagerII1931} with a hydrodynamic description of the long-time evolution of the long-wavelength components of the energy-density fluctuations, the small-frequency/small-wavevector limit of the EDSF can be shown to read:%\cite{KADANOFF1963419,forster,hansen2013}
%\resubcancel{The hydrodynamic (low-frequency, long-wavelength) limit of the EDSF reads:}
\cite{forster,hansen2013}
\begin{equation}
    \widetilde C(\kk,\omega)=k_BT^2\frac{k^2D_Tmnc_p}{\omega^2+k^4D_T^2}, \label{eq:C_hydro} 
\end{equation}
%
%\resubcancel{The equivalence between Eqs. \eqref{eq:GK} and \eqref{eq:GK_chi} follows immediately from Eq. \eqref{eq:C_hydro}, given the relation between density and longitudinal-current correlation functions, due to the effectiveness of the continuity equation: $\widetilde C_{\epsilon\epsilon}(\kk,\omega)=\frac{k^2}{\omega^2}C_{jj}(\kk,\omega)$.}
 Inspection of Eq. \eqref{eq:C_hydro} reveals that for large wavelengths the static limit of the EDSF reads:
\begin{equation}
\widetilde C(\kk,0)=k_BT^2\frac{mnc_p}{k^2D_T}.
\end{equation}
We conclude that the heat conductivity can be equivalently expressed as:

\begin{align}
    \kappa &=\lim_{\kk\to0} \kappa(\kk),\label{eq:palmer} \\
    \kappa(\bm k) &= \frac{k_B T^2}{k^2 } \frac{(mnc_p)^2}{
    % \lim_{\omega\to 0} 
    \widetilde C(\kk,\omega=0)}.  \label{eq:k palmer}  
\end{align}
Note the non-commutativity of the limits in Eq. \eqref{eq:GK_chi}, manifested by Eq. \eqref{eq:C_hydro}.  This well-known singularity, which has been extensively discussed in the literature following Kadanoff and Martin's landmark paper,\cite{KADANOFF1963419} should not come as a surprise, for the zero-frequency limit of a generic response function is a static susceptibility. It is an obvious consequence of time-reversal invariance that, strictly speaking, a static perturbation cannot induce a stationary current. The interpretation of this equation when the limits are performed in the correct order is as follows: While the long-time limit of the current induced by a temperature fluctuation vanishes at all finite wavelengths, as the system approaches equilibrium, at any finite time, the current tends to a finite, time-independent, value as the wavelength of the perturbation grows large. This behavior defines a stationary current in response to the temperature fluctuations.

%in Appendix \ref{app:GaugeInvariance}.

\subsection{Gauge invariance} \label{app:GaugeInvariance}
The energy density, $\widehat e(\rr)$, entering Eq. \eqref{eq:Ctilde} is intrinsically ill-defined, as it is affected by a so-called \emph{gauge freedom}, deriving from the insensitivity of the total energy---and therefore of all the macroscopic thermal properties---on the addition of the divergence of a bounded vector field to the energy density: $e(\rr)\to e(\rr)+ \grad\cdot \bm p(\rr)$. The independence of the heat conductivity on the specific representation of the energy density is discussed, e.g., in Refs. \cite{marcolongo2016,grasselli2021,Ercole2017, marcolongo_2020}, and demonstrated below in the context of the present approach.

In order to prove the gauge invariance of Eq. \eqref{eq:palmer}, we examine how a gauge transformation, $e^\prime(\rr)= e(\rr)+\nabla\cdot \bm p (\rr)$, affects the zero-frequency value of the EDSF, $\widetilde{C}(\kk,\omega=0)$. Under such a transformation, the integrand in Eq. \eqref{eq:Ctilde} would go into:
\begin{widetext}
\begin{equation}
    \left\langle \widehat{\widetilde e^\prime}(\kk,t) \widehat{\widetilde e^\prime}(-\kk,0)\right\rangle = \left\langle \widehat{\widetilde e}(\kk,t) \widehat{\widetilde e}(-\kk,0)\right\rangle + \left\langle \bigl ( \kk\cdot \widehat{\widetilde{\mathbf{p}}}(\kk,t) \bigr ) \bigl ( \kk \cdot \widehat{\widetilde{\mathbf{p}}}(-\kk,0) \bigr ) \right\rangle + 2\kk\cdot\Im\left\langle \widehat{\widetilde e}(\kk,t) \widehat{\widetilde{\mathbf{p}}}(-\kk,0)\right\rangle. \label{eq:c_gauge}
\end{equation}
%\end{widetext}
%
According to the Wiener-Kintchine theorem \cite{wiener1930generalized,Khintchine1934}, the Fourier transform of the time auto-correlation function of a zero-mean stationary stochastic process, $\widehat X(t)$, in the large-time limit is proportional to the variance of the Fourier transform of the process, $\widehat{\widetilde X}_\tau (\omega)=\int_{-\frac{\tau}{2}}^{\frac{\tau}{2}} \widehat X(t)e^{i\omega t}dt$: 
\begin{equation}
\int_{-\infty}^{\infty} \langle \widehat X(t) \widehat X(0)  \rangle e^{i\omega t} dt = \lim_{\tau\to\infty} \frac{1}{\tau} \left | \widehat{\widetilde X}_\tau(\omega)\right |^2.
\end{equation}
The $\omega=0$ limit of this relation reads:
\begin{equation}
\int_0^\infty \langle \widehat X(t) \widehat X(0)  \rangle dt = \lim_{\tau\to\infty} \frac{1}{2\tau} \left | \int_0^\tau\widehat{X}(t)dt \right |^2. \label{eq:EinsteinHelfand}
\end{equation}
When $\widehat X(t)$ is a conserved flux, the left-hand side of Eq. \eqref{eq:EinsteinHelfand} is the Green-Kubo expression of a transport coefficient. This equation is a generalization of Einstein's celebrated relation between a particle's diffusivity and its velocity auto-correlation function.\cite{Einstein:1905} Einstein's formula was extended to generic transport coefficients by E. Helfand in 1960 \cite{Helfand1960}.

According to these considerations, the $\omega=0$ value of the transformed EDSF can be computed as
%\begin{widetext}
\[
\widetilde{C}^\prime(\kk,0) = \lim_{\tau\to\infty}\frac{V}{2\tau} \left [ \left\langle\abs{\int_{0}^{\tau} \widehat{\widetilde{e}}(\kk,t) \, dt}^2\right\rangle + \left\langle\abs{\int_{0}^{\tau} \bm{k}\cdot\widehat{\widetilde{\mathbf{p}}}(\kk,t) \, dt}^2\right\rangle + 2  \bm{k}\cdot\Im\left\langle\int_{0}^{\tau}\widehat{\widetilde{e}}(\kk,t) \, dt\int_{0}^{\tau} \widehat{\widetilde{\mathbf{p}}}(-\kk,t^\prime) \, dt^\prime \right\rangle \right ].\label{eq:gauge change}
\]
\end{widetext}
The second term of Eq. \eqref{eq:gauge change} is bounded from above by 
\begin{equation}
    \lim_{\tau\to\infty}\frac{V}{2\tau}\left\langle\int_{0}^{\tau} \abs{\bm{k}\cdot\widehat{\widetilde{\mathbf{p}}}(\kk,t)}^2 \, dt\right\rangle= \frac{V}{2}k^2\left\langle \abs{\widehat{\widetilde{\mathbf{p}}}(0)}^2\right\rangle + {\mathcal O}(k^4),
\end{equation} 
and therefore, since $\mathbf{p}(\rr)$ has a finite variance, it vanishes in the $\kk\to0$ limit. Since in $\kk\to0$ the limit both $\widehat{\widetilde{e}}(\kk,t)$ and $\widehat{\widetilde{\mathbf{p}}}(-\kk,t^\prime)$ are real, the third term in Eq. \eqref{eq:gauge change} vanishes as well. 
% Thus, Eq. \eqref{eq:GK}, and the expression via the energy density, Eq. \eqref{eq:palmer}, are proper estimators of $\kappa$. 
While at finite $\kk$ it is to be expected that $\kappa(\kk)$ in Eq. \eqref{eq:k palmer} may be affected by the specific choices of $\bm p(\rr)$, its long-wavelength limit does not depend on it, thus providing an independent demonstration of the gauge invariance of the heat conductivity, particularly adapted to the present framework.

\section{Computer simulations} \label{sec:ComputerSimulations}
The method embodied in Eq. \eqref{eq:palmer} has been benchmarked against standard GK calculations based on Eq. \eqref{eq:GK} in the cases of liquid water, liquid Argon and solid amorphous Silica. Water was modeled with a sample of 216 molecules interacting through the SPC/E force field \cite{spce} at $\mathrm{T=300~K}$ and $p=0~\mathrm{bar}$. Argon was modeled with a sample of 864 atoms interacting through a Lennard-Jones potential \cite{ROWLEY1975401} at $\mathrm{T=100~K}$ and $p=0~\mathrm{bar}$. Amorphous Silica was modeled with a sample of 648 atoms interacting via the BKS potential \cite{BKS1990} at $500~\mathrm{K}$. Long EMD simulations were run in the micro-canonical (NVE) ensemble after careful equilibration performed in a number of different ensembles. In all cases periodic boundary conditions (PBC) were used. All the technical details of the simulations are reported in Appendix \ref{app:CompDetails}. 
In order to estimate the $\omega= 0$ value of the ESDF in Eq. \eqref{eq:k palmer}, the time series of the Fourier transform of the energy density, $\tilde e(\kk_\ell,t)=\sum_n \epsilon_n(t)e^{-i\kk_\ell\cdot \RR_n(t)}$, was first collected along the EMD trajectory at regular time steps. In the preceding expression $\epsilon_n(t)$ and $\RR_n(t)$ are the local energy and position of $n$-th atom, respectively, at time $t$. Leveraging the Wiener-Kintchine theorem, the ESDF is then evaluated as:
\begin{equation}\label{eq:Wiener-Khintchine}
    \widehat{\widetilde C}(\kk,\omega) \sim \frac{2}{\tau}\left | \int_0^\tau \widehat{\widetilde e}(\kk,t)e^{i\omega t}dt \right |^2,
\end{equation}
where $\tau$ is the length of the EMD run and the integral is actually evaluated as a discrete Fourier transform, as explained, \emph{e.g.}, in Refs. \onlinecite{Ercole2017} and \onlinecite{sportran}. In the signal-processing literature, a sample of a power spectral density (PSD) drawn from a finite and discrete time series is often referred to as a \emph{periodogram}. In Eq. \eqref{eq:Wiener-Khintchine} and in the rest of this section a hat indicates that the quantity it is put on (the periodogram and the energy density, in the present case) is actually a sample drawn from a stochastic process. It can be demonstrated that the periodogram constitutes a set of stochastic variables that are uncorrelated for different frequencies and that, for a given frequency, $\omega_j=\frac{2\pi j}{\tau}$, are distributed as:
\begin{equation} \label{eq:periodogram}
    \widehat{\widetilde C}(\kk,\omega_j)= \frac{1}{2} \widetilde C(\kk,\omega_j) \widehat\xi_j,
\end{equation}
where the $\widehat \xi$'s are independent $\chi^2_2$ variates. From Eq. \eqref{eq:periodogram} it follows that the (energy-density) periodogram is an unbiased estimator of the PSD (the ESDF, in this case), $\bigl \langle \widehat{\widetilde C}(\kk,\omega_j) \bigr \rangle = \widetilde C(\kk,\omega_j)$. This estimator is not consistent, though, because the variance of any of its values is independent of the length of the EMD run. In order to get such a consistent estimator (at the price of introducing some bias), it is expedient to make use of the so called \emph{cepstral analysis} method \cite{Bogert1963,Ercole2017,sportran}. The \emph{cepstrum} is defined as the (inverse) discrete Fourier transform of the logarithm of the periodogram. The logarithm turns the multiplicative noise affecting the periodogram into an additive noise affecting the cepstrum; furthermore, the PSD, as well as its logarithm, are much smoother than the periodogram and the cepstrum itself. Therefore, a consistent estimate of the log-PSD can be obtained by applying a low-pass filter to the cepstrum. The optimal value of the low-pass \emph{quefrency} \cite{Bogert1963}---\emph{i.e.} the number of inverse Fourier coefficients of the cepstrum---is determined in such a way that the estimated bias introduced by the filter is of the order of the resulting statistical noise. This number can be obtained by any model-selection method, and in our applications we use the \emph{Akaike information criterion} \cite{Akaike1974,Ercole2017, sportran}. 

\begin{figure}
    \centering
    \includegraphics[width=\linewidth]{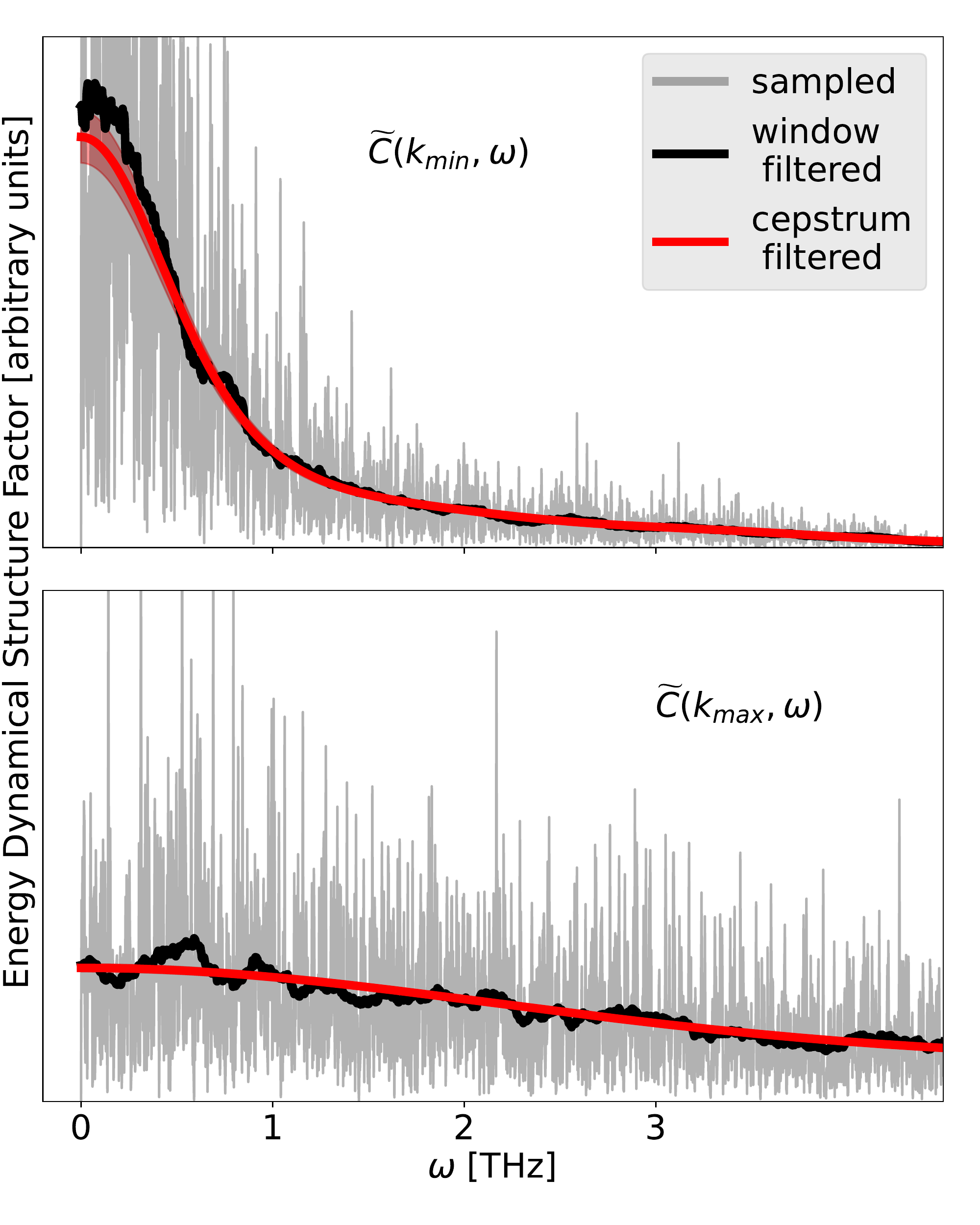}
   \caption{Upper panel: EDSF of liquid Argon computed at PBC-compatible $\kk$-vectors of minimum magnitude, $k_{min}=1.77\,\mathrm{nm}^{-1}$. Lower panel: same, but for the $\kk$-vectors, $k_{max}=11.3\,\mathrm{nm}^{-1}$.  The \emph{window-filtered} line refers to a moving average\cite{MovingAverage} performed width a window with of $\Delta=0.2~\mathrm{THz}$. The shaded red area indicates the statistical incertitude.}
    \label{fig:cepstral}
\end{figure}

With the aim of evaluating Eq. \eqref{eq:palmer}, the EDSF was first sampled on a homogeneous grid of wave-vectors, $\{\kk_\ell\}$, with magnitude smaller than a preassigned cutoff, $|\kk_\ell|\le k_{max}$, compatible with the chosen PBCs. Because of spherical symmetry, the EDSF only depends on the modulus of its wave-vector argument and can therefore be averaged accordingly. In practice, PBCs reduce the symmetry from spherical to cubic, thus making some PBC-allowed wave-vectors non-equivalent to others with the same magnitude. This artifact is eliminated by averaging the estimates of the EDSF corresponding to different wave-vectors of the same magnitude, thus further reducing the variance of any individual value. The set of wave-vectors with a same magnitude, $k_i$, will be referred to as the $k_i$-star of wave-vectors. In Fig. \ref{fig:cepstral} we display the EDSF of liquid Argon at the the $k$-star of minimum magnitude compatible with the chosen PBCs, $k_{min}=1.77\,\mathrm{nm}^{-1}$, and the one of largest magnitude utilized in our Bayesian analysis (see below), $k_{max}=11.3\,\mathrm{nm}^{-1}$. The increasing sharpness of the peaks for increasing wavelength (decreasing wave-vector) is a signature of their hydrodynamic nature, as manifested by Eq. \eqref{eq:C_hydro}, and makes cepstral analysis increasingly difficult to perform, thus giving rise to a larger statistical incertitude.
%Since during the NVE simulation energy is a conserved quantity, the EDSF related to the $\kk_{min}$ has a 0-frequency peak much more narrow than the one associated to $\kk_{max}$. Owing to the presence of this feature, the cepstral analysis may need many cepstral coefficients and thus the spectrum would be  challenging to analyze \cite{Ercole2017}.
Because of isotropy, the wave-vector dependence of the conductivity, Eq.  \eqref{eq:k palmer}, is an even function of the squared modulus of $\kk$, which we assume to be well approximated by a polynomial of order $M$:
\[\kappa(\kk)\approx w_0+w_1k^2+w_2k^4\dots+w_Mk^{2M}\label{eq:polynomial_fit}.\]
In order to determine the coefficients of the polynomial fit in Eq. \eqref{eq:polynomial_fit}, $\bm{w}=\{w_0\cdots w_M\}$, we relied on a Bayesian inference method, by seeking to maximize their posterior probability, given the occurrence of the measured data, $\mathcal D$: $p\left(\bm{w}\vert \mathcal{D}\right)=p\left(\mathcal{D}\vert \bm{w} \right)p(\bm{w}) / p(\mathcal{D}) $, where $p(\mathcal{D}) $ and $p(\bm{w})$ are the data and model-parameter priors, respectively. The data are the conductivities measured at each $k$-star, which are independent normal variates whose expectations, $\kappa_i$, and variances, $\sigma_i^2$ are estimated by the cepstral analysis procedure outlined above. The normality of the distribution of the measured values of $\kappa(\kk)$ is showcased in Appendix \ref{app:normality}. The probability of the data conditional to the values of the coefficients of the polynomial is thus given by $p\left(\mathcal{D}\vert \bm{w} \right) \propto e^{-\mathcal{L}(\bm{w})}$, where $\mathcal{L}(\bm{w})=\sum_i \frac{\left(\kappa_i - \bm{w}\cdot\bm{\Phi}_i\right)^2}{2\sigma_i^2}$, is the likelihood function and $\bm{\Phi}_i$ is an array of monomials of even degree up to order $2M$, evaluated at the $k$-stars, $k_i$, for which the $\kk$-dependent conductivity has been estimated: $\bm{\Phi}_i=\bm \Phi(k_i)\equiv \{1,k_i^2\dots k_i^{2M}\}$.
%, computed at the wave-vectors being sampled, and $\sigma_i$ is the standard deviation of $\kappa_i$, as estimated via block analysis. 
%In order to compute the probability, $p\left( \bm{w}\vert\mathcal{D}\right)$, that $\bm{w}\cdot \bm{\Phi}(\kk)$ is the correct function from which the dataset, $\mathcal{D}\equiv\{\kappa_i\}$, was sampled, we assume that the probability of the dataset, given the parameters, is $p\left(\mathcal{D}\vert \bm{w}\right)\propto \exp\left [- \mathcal{L}(\bm{w}) \right ]$ (see Appendix \ref{normality}). 
% Thus, the posterior distribution function is determined thanks to the Bayes theorem, $p\left(\bm{w}\vert \mathcal{D}\right)p(\mathcal{D})=p\left(\mathcal{D}\vert \bm{w} \right)p(\bm{w})$.
%The probability that the data-set, $\mathcal{D}\equiv\{\kappa_i\}$, is generated from the function $\bm{w}\cdot \bm{\Phi}_i$ is proportional to the exponential of the negative of the likelihood, $p\left(\mathcal{D}\vert \bm{w}\right)\propto \exp\left [- \mathcal{L}(\bm{w}) \right ]$. The posterior distribution, $p\left( \bm{w}\vert\mathcal{D}\right)$, i.e. the probability that $\bm{w}\cdot \bm{\Phi}(\kk)$ is the correct function from which  $\mathcal{D}$ was sampled, is determined thanks to the Bayes theorem: $p\left(\bm{w}\vert \mathcal{D}\right)p(\mathcal{D})=p\left(\mathcal{D}\vert \bm{w} \right)p(\bm{w})$.
We now make the further assumption that the prior distribution of the parameters is normal, $p(\bm w)=\left(\frac{\alpha}{2\pi}\right)^{\frac{M}{2}}\exp\left [-\alpha\norm{\bm{w}}^2 \right ]$, thus allowing the extrapolation to prevent over-fitting, which, in ordinary linear regression, would be equivalent to introducing a regularization term, $\alpha\norm{\bm{w}}^2$.
%Assuming that the prior distribution is normal, $p(\bm w)=\left(\frac{\alpha}{2\pi}\right)^{\frac{M}{2}}\exp\left [-\alpha\norm{\bm{w}}^2 \right ]$, is equivalent to introducing a  regularization term, $\alpha\norm{\bm{w}}^2$, in ordinary linear regression, so as to prevent over-fitting. The prior distribution of the parameters depends implicitly on their number, $M+1$, and on the hyper-parameter $\alpha$.
By leveraging again Bayes theorem, the optimal number of parameters, $M$, and regularization parameter, $\alpha$, are determined as those that maximize the probability of their occurrence, conditionally to actual observation of the data set:
\begin{equation}
    p(M,\alpha\vert \mathcal D)\propto p(\mathcal D|M,\alpha) p(M,\alpha).
\end{equation}
The actual procedure that we followed to determine $M$, $\alpha$, $\bm w$, and, therefore, $\kappa=w_0$ is described in  Refs. \onlinecite{drigo2023jctc,bishop2006}.

\begin{figure}
    \centering
    \includegraphics[width=\linewidth]{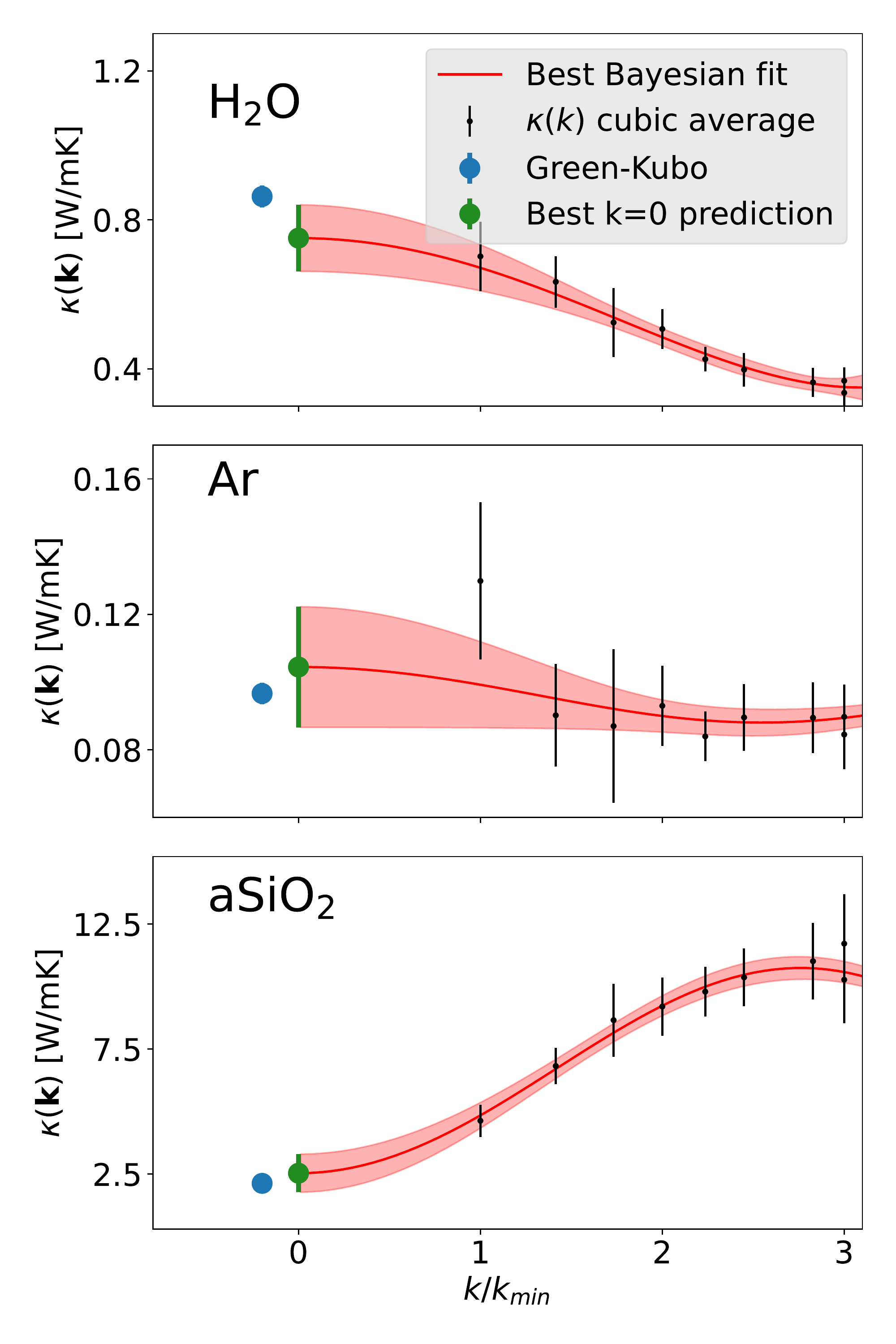}
    \caption{Upper panel: Wave-vector dependence of the heat conductivity, $\kappa(\kk)$, Eq. \eqref{eq:k palmer}, of SPC/E water at $T=300~\mathrm{K}$ and $0~\mathrm{bar}$, as evaluated from the EMD simulation, $k_{min}=3.40\,\mathrm{nm}^{-1}$. Central panel: same for liquid Argon at $T=100~\mathrm{K}$ and $0~\mathrm{bar}$, $k_{min}=1.77\,\mathrm{nm}^{-1}$. Lower panel: same for solid amorphous Silica at $T=500~\mathrm{K}$, $k_{min}=3.04\,\mathrm{nm}^{-1}$. The data are averages of $\kk$-vectors with a same magnitude, compatible with the PBCs in use. \emph{Cubic average} refers to an average performed over $\kk$-vectors that are strictly equivalent by cubic symmetry, i.e. excluding vectors that are equivalent by spherical, but not cubic symmetry. The red line indicates the Bayesian fit with the associated estimated uncertainty. At $\kk=0$ we report the Bayesian extrapolation and the GK estimate, on the same EMD trajectory, of the heat conductivity. } \label{fit}
\end{figure}
In Fig. \ref{fit} we display the wave-vector dependence of Eq. \eqref{eq:k palmer}, $\kappa(\kk)$, computed for SPC/E water, liquid Argon and solid amorphous Silica via the cepstral analysis, and the $\kk\to0$ extrapolation of Eq. \eqref{eq:palmer}, $\kappa$, obtained from the Bayesian regression method. The results are in good accordance with the GK integral estimated via the cepstral analysis on the same trajectory, also reported in the figures. 
In Appendix \ref{app:size-convergence} we study the finite size effects on the computation of Eq. \eqref{eq:k palmer}. We studied the size convergence of Eq. \eqref{eq:k palmer} for SPC/E water modeled with samples of different linear box size, $L$: $18.5~\mathrm{\AA}$, $31.1~\mathrm{\AA}$ and $37.4~\mathrm{\AA}$; corresponding to $216$, $1000$ and $1728$ molecules respectively.

\begin{figure}
    \centering
    \includegraphics[width=\linewidth]{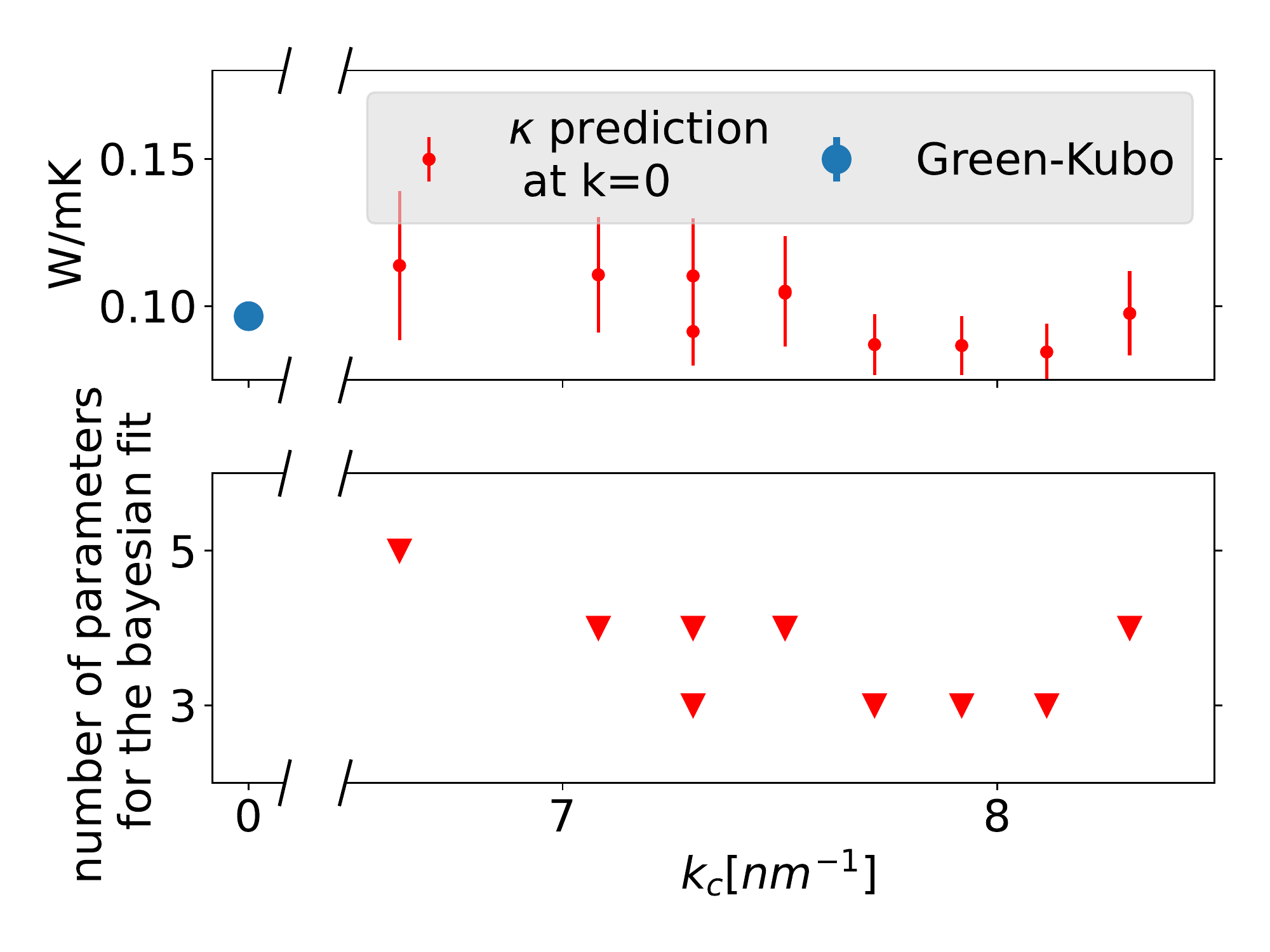}
    \caption{Upper panel: Bayesian prediction of liquid Argon at $100 K$ and $0$ bar, at $\kk=0$ as a function of the maximum $\kk$-vector included in the fit, $k_c$, confronted with the GK integral computed on the same trajectory with the cepstral analysis. Lower panel: Number of parameters of the Bayesian regression fit as a function of the maximum $\kk$-vector included in the fit. } \label{bayespredargon}
\end{figure}

In Fig. \ref{bayespredargon} we display the Bayesian extrapolation ($\kk=0$) of the heat conductivity of liquid Argon as a function of the maximum magnitude of the wave-vectors considered in the fit, $k_c$. The Bayesian extrapolation is robust and converges to the GK result computed on the same trajectory, also reported in the figures, when we enlarge the data set, thus demonstrating the numerical stability and statistical consistency of our method.

%\begin{figure}
%    \centering
%    \includegraphics[width=\linewidth]{2161nsspce/convergence.pdf}
%    \caption{Upper panel: Bayesian prediction of SPC/E water at $\kk=0$ as a function of the maximum $\kk$-vector included in the fit confronted with the GK integral computed on the same trajectory with the cepstral analysis. Lower panel: Number of parameters of the Bayesian regression fit as a function of the maximum $\kk$-vector included in the fit.}
%    \label{bayespred}
%\end{figure}

\section{Conclusions}

In this work we have integrated a method originally proposed by Palmer in 1994,\cite{palmer1994} to compute the heat conductivity in extended systems from energy-density fluctuations, with modern data and signal analysis techniques that considerably enhance its applicability. As a downside, %of the presented procedure, 
as already anticipated by Palmer, the numerical efficiency of the methodology presented here appears to be poorer than that of the standard GK procedure based on the fluctuations of the energy flux. 
In the specific case of Argon, we estimated that, in order to achieve the same statistical accuracy of the standard GK procedure, 
% the cepstral estimate of the GK integral, 
one would need EMD trajectories one-order of magnitude as long. The enhanced statistical performance of current-based methods for estimating transport coefficients, as opposed to density-based methods, can be ascribed to the necessity of performing long-wavelength extrapolations when utilizing the latter. Avoiding any such extrapolation would significantly enhance the performance of any statistical analysis, no matter how smart this may be. Of course, this numerical superiority may, or may not, be tarnished by the need to devise and implement an explicit expression for the macroscopic energy flux. 

In particular, it may be challenging, if not impossible, to devise an analytical form of the energy flux when advanced energy functionals, such as hybrid or meta-GGA, are used for \emph{ab initio} MD. We believe that a density-based approach, such as the one examined in the present paper or in CF's \cite{cheng2020}, would be mandatory in these cases. The relevance of these cases is however increasingly limited by the emergence of artificial-intelligence methods to train classical force fields with quantum mechanical accuracy \cite{Mishin2021}, which allow for a fast and accurate evaluation of various transport coefficients using the standard GK approach, including the heat conductivity. \cite{tisi2021,Malosso2022} Furthermore, all the density-based methods proposed so far are at present limited to one-component systems, whereas a systematic extension of the standard GK one has been recently established \cite{Bertossa2019}. In summary, we conclude that standard current-based methods are preferable in all those cases where an explicit expression for the energy flux is available. In the rare cases where it is not (e.g. for \emph{ab initio} MD with advanced functionals), density-based methods may be the way to go. The method discussed here and the one proposed by CF \cite{cheng2020} are conceptually similar, but the latter is in practice only applicable when the Landau-Planczek ratio is sizeable, i.e. for compressible fluids, whereas the former can be applied to solids and incompressible fluids as well. We also hint at the superiority of the data analysis technique introduced in Ref. \onlinecite{drigo2023jctc} and utilized here, which allows for an easy control and optimal estimate of the statistical accuracy of the transport coefficients. This technique would readily integrate into CF's approach \cite{cheng2020}, as well as in any methodology relying on the long-wavelength extrapolation of correlation functions estimated for finite, periodic models.

\section*{Data Availability}

The data and scripts that support the plots and relevant results within this paper are available on the Materials Cloud platform~\cite{materialscloud}. See DOI:XXX. 

\begin{acknowledgments}
The authors are grateful to Riccardo Bertossa, Alfredo Fiorentino, Federico Grasselli, Paolo Pegolo, and Davide Tisi for many insightful suggestions and valuable discussions. This work was partially supported by the European Commission through the \textsc{MaX} Centre of Excellence for supercomputing applications (grant number 101093374) and by the Italian MUR, through the PRIN project \emph{FERMAT} (grant number 2017KFY7XF) and the Italian National Centre from HPC, Big Data, and Quantum Computing (grant number CN00000013).
\end{acknowledgments}

\section*{Competing interests}
The authors have no conflicts to disclose.

\section*{Author contributions}
ED, MGI, and SB conceived the study and provided the overall conceptual framework. ED and MGI conducted the analytical work and developed the necessary computer codes. ED run and analyzed the computer simulations.
SB provided supervision and guidance throughout the project. The manuscript was collaboratively written by ED and SB.

% \section*{Data availability}
% Numerical data supporting the plots and relevant results within this paper are available on the Materials Cloud Platform \cite{materialscloud}.

\appendix

\begin{figure}
    \centering
    \includegraphics[width=\linewidth]{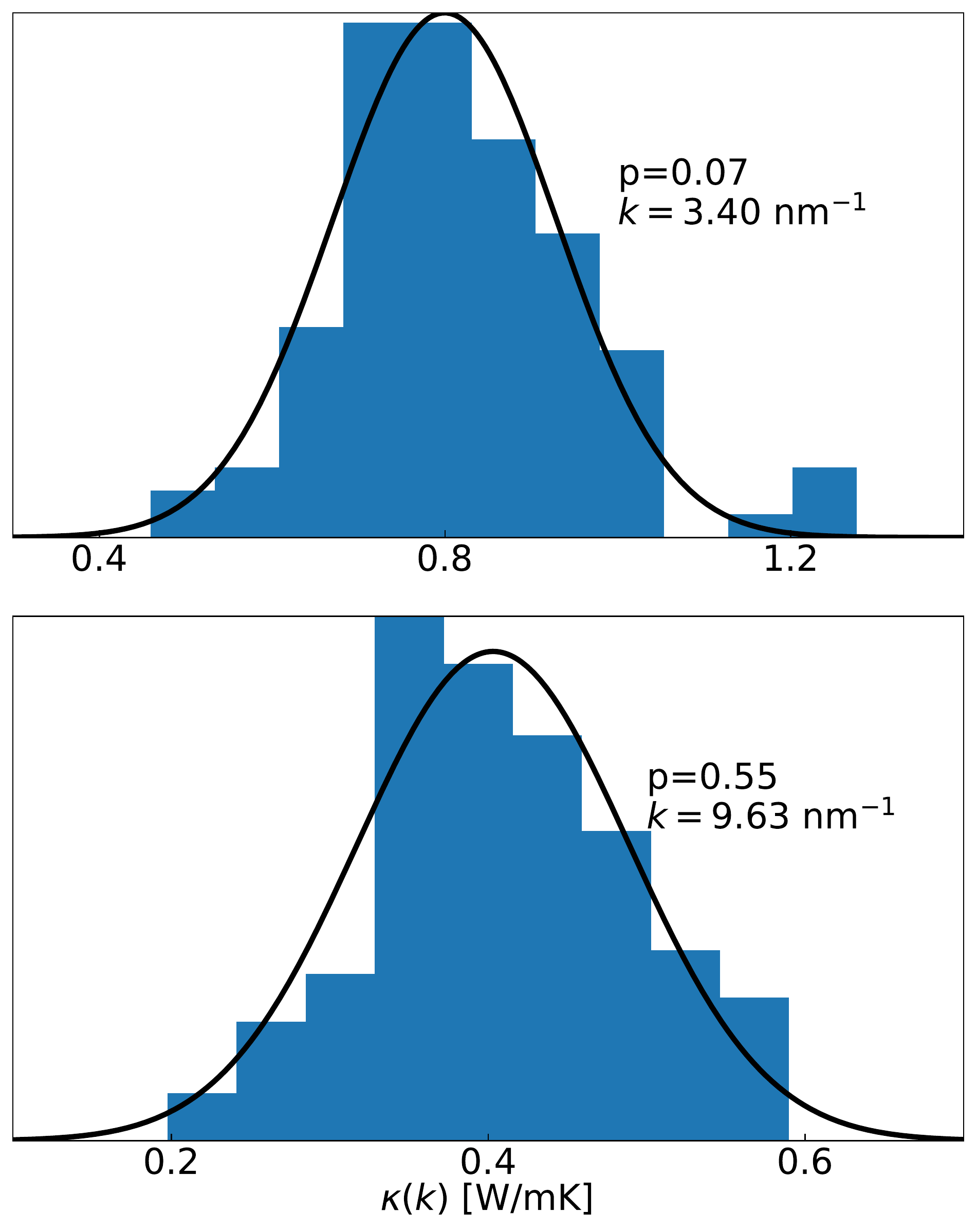}
    \caption{Distribution of the wave-vector dependence of the heat conductivity, $\kappa(\kk)$, Eq. \eqref{eq:k palmer}, of SPC/E water at $T=300K$ and $0$ bar as evaluated from multiple $200 ps$-long EMD segments extracted from a $20 ns$ long trajectory and the p-value of the Shapiro-Wilk normality test. The solid lines represent a Gaussian curve fitted on the distribution of the heat conductivities at the wavevector presented.}
    \label{gauss}
\end{figure}

\section{Computational details} \label{app:CompDetails}
We modeled water with the SPC/E force field that assumes the intra-molecular bond lengths and angles to be constrained to their equilibrium values, while inter-molecular atomic forces are described by Coulomb-plus-Lennard-Jones potentials whose parameters are fitted to experiment \cite{spce}. The resulting agreement between computed and measured static and dynamic properties, such as the radial distribution function ad diffusivity, is very good. The long-range Coulomb interaction was treated via the Ewald summation method with a $9$ \AA\, cutoff. We have performed NVE EMD simulations of $216$ SPC/E water molecules at $T=300$ K and $p=0$ bar, using PBCs. After a careful equilibration in the NPT and NVT ensembles for $125$ ps, we sampled the energy density from a $1$-ns NVE trajectory  with a time step of $0.25$ fs. The energy density was computed and dumped at intervals of $2.5$ fs.
%We tested different models of water such as the SPC-Flex \cite{spcfw} and the TIP4P/Ice \cite{tip4pice} with the same simulation protocol as for the SPC/E model.
Liquid Argon was simulated at $T=100$ K and $p=0$ bar using a sample of 864 atoms interacting through an iter-atomic Lennard-Jones potential whose parameters are fitted to experiments \cite{ROWLEY1975401}. After a preliminary equilibration in the NPT ensemble, we sampled the system from a $2$-ns NVE trajectory using a time step of $1$ fs. In this case, the energy density was computed and dumped at intervals of $5$ fs. 
Solid amorphous Silica was simulated at $500~\mathrm{K}$ using a sample of 648 atoms modeled via the BKS potential\cite{BKS1990}. The initial configuration was taken from Ref.~\onlinecite{Fiorentino2023} and optimized  so as to make atomic forces smaller than a preassigned threshold of $10^{-8}~\mathrm{eV/\AA}$. After careful equilibration for $125~\mathrm{ps}$ in the NVT ensemble, we simulated the system for $1~\mathrm{ns}$ in the NVE ensemble with a timestep of $0.25~\mathrm{fs}$. The energy density was computed and dumped at intervals of $5$ fs. 
All simulations were performed with the parallel code \texttt{LAMMPS} \cite{Lammps, lammps1} using the velocity-Verlet algorithm.

\section{Statistical properties of the dataset} \label{app:normality}
Our Bayesian regression procedure is based on the assumption that the data (the values of the wavevector-dependent heat conductivities, $\kappa(\kk)$, Eq. \eqref{eq:k palmer}) are normal variates at any value of their argument, $\kk$.
% In the Bayesian regression fit we chose the likelihood function to be $\mathcal{L}(\bm{w})=\sum_i \frac{\left(\kappa_i - \bm{w}\cdot\bm{\Phi}_i\right)^2}{2\sigma_i^2}$, where $\kappa_i$ is our EMD estimate of the heat conductivity at wave-vector $\kk_i$, $\bm{\Phi}_i$ is the basis set  consisting of  monomials of even degree, $\bm{\Phi}_i=\bm \Phi(\kk_i)\equiv \{1,k_i^2\dots k_i^{2M}\}$, computed at the wave-vectors being sampled, and $\sigma_i$ is the standard deviation of $\kappa_i$, as estimated via the cepstral analysis. By doing so, we implicitly assumed that  the dataset, $\mathcal{D}=\{\kappa_i\}$, is normally distributed, and its distribution function,  $p\left(\mathcal{D}\vert \bm{w}\right)$, is $p\left(\mathcal{D}\vert \bm{w}\right)\propto \exp\left [- \mathcal{L}(\bm{w}) \right ]$. 
In order to evaluate the soundness of this hypothesis, we have examined the statistics of the values of the $\kappa$'s collected from $200$-ps segments extracted from our $20$-ns long EMD trajectory for liquid water. In Fig. \ref{gauss} we display the distribution of the data thus collected, corresponding to the wavevectors of minimum magnitute, $k_{min}=3.40\,\mathrm{nm}^{-1}$ and of maximum magnitude, $k_{max}=9.63\,\mathrm{nm}^{-1}$ utilized in our Bayesian procedure. In both cases, the data pass the Shapiro-Wilk (SW) normality test \cite{shapirowilk, Malosso2022} with respect to a standard significance level of $\alpha=0.05$.

\section{Finite-size effects} \label{app:size-convergence}

In order to study the size-convergence of the method, we computed the wave-vector dependence of the heat conductivity, $\kappa(\kk)$, Eq. \eqref{eq:k palmer}, for SPC/E water at $300~\mathrm{K}$ and $0~\mathrm{bar}$, for three different linear box sizes, $L$: $18.5~\mathrm{\AA}$, $31.1~\mathrm{\AA}$ and $37.4~\mathrm{\AA}$; corresponding to $216$, $1000$ and $1728$ water molecules respectively. Appendix \ref{app:CompDetails} describes the simulation protocol that we employed for the simulations of SPC/E water at each size.  

\begin{figure}
    \centering
    \includegraphics[width=\linewidth]{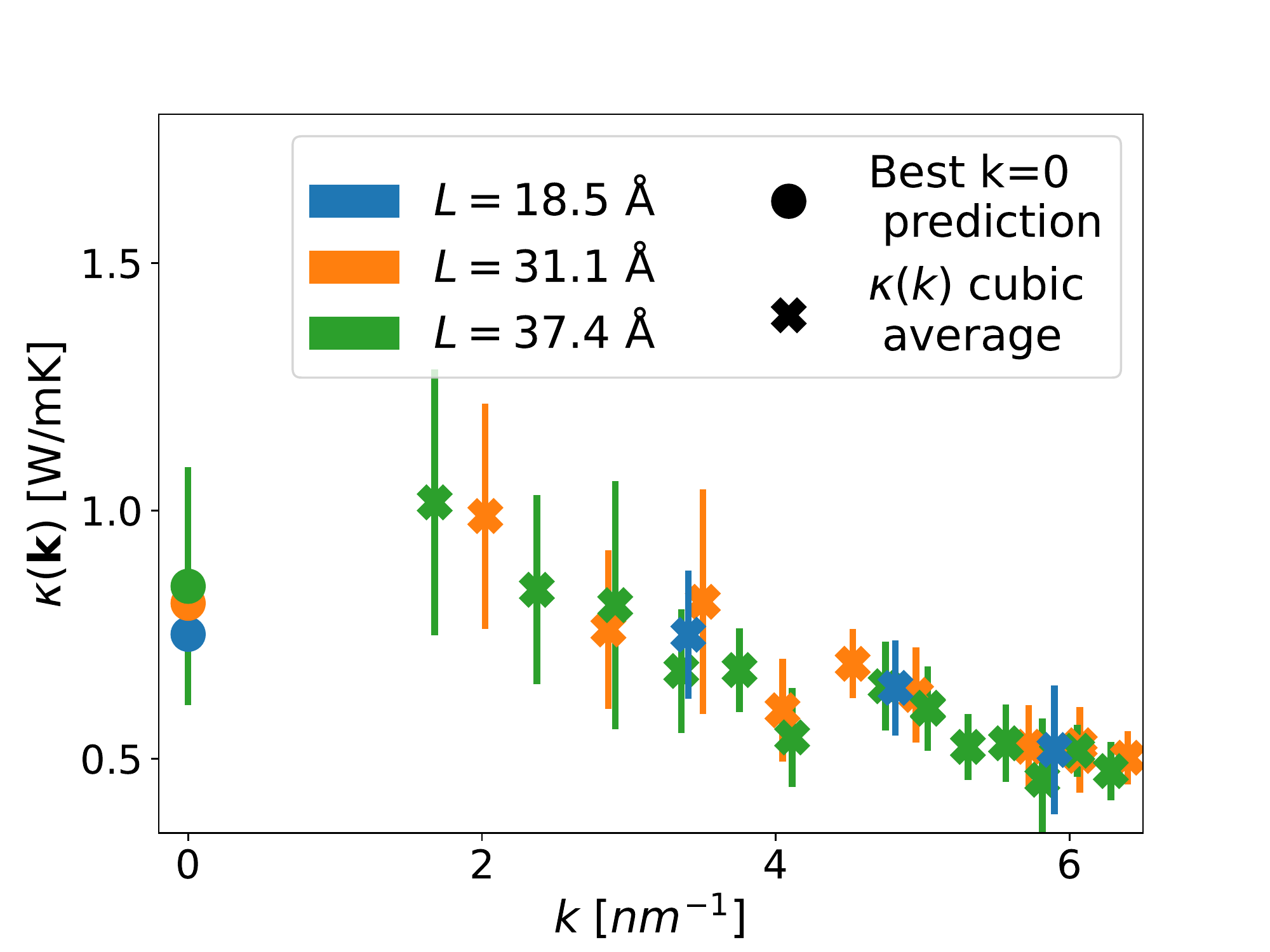}
    \caption{Wave-vector dependence of the heat conductivity, $\kappa(\kk)$, Eq. \eqref{eq:k palmer}, of SPC/E water at $300~\mathrm{K}$ and $0~\mathrm{bar}$, as evaluated from the EMD simulations for three different linear box sizes, $L$: $18.5~\mathrm{\AA}$, $31.1~\mathrm{\AA}$ and $37.4~\mathrm{\AA}$; corresponding to $216$,  $1000$ and $1728$ water molecules respectively. The data are averages of $\kk$-vectors with a same magnitude, compatible with the PBCs in use. \emph{Cubic averages}, represented as crosses, refer to the averages performed over $\kk$-vectors that are strictly equivalent by cubic symmetry, i.e. excluding vectors that are equivalent by spherical, but not cubic symmetry. At $\kk=0$ we report, as dots, the Bayesian extrapolations of the heat conductivity at each size.}
    \label{size-convergence}
\end{figure}

In Fig. \ref{size-convergence} we display the size-convergence of  Eq. \eqref{eq:k palmer}, for SPC/E water. Increasing the number of molecules, e.g. augmenting the box size at fixed mass density, allows to investigate further the $\kk\to0$ limit of $\kappa(\kk)$, computing smaller module wave-vectors. Focusing on the wave-vectors that are compatible with all the sizes we considered, e.g. approximatively $3.40~\mathrm{nm^{-1}}$, compared to the simulation of $1728$ molecules, already with $216$ molecules we were able to obtain a well converged estimate of Eq. \eqref{eq:k palmer}.

\section*{references}
\bibliography{biblio}% Produces the bibliography via BibTeX.
\end{document}